# Scaling law in thermal phenomena

Janina Marciak-Kozlowska
Institute of Electron Technology,Al. Lotnikow 32/46,
02-668 Warsaw, Poland
and
Miroslaw Kozlowski

Abstract
In this paper the scaling law for the relaxation times in thermal phenomena is investigated. It is shown that dependent on the value of the parameter $K=E/m(c\alpha)^2$, where $E$ is the energy which is delivered to the system , $m$ is the *parton* mass and $\alpha=1/137$ for electromagnetic interaction and $\alpha =0.16$ for strong interaction respectively, heat transport is diffusive, for $K<1$, or contains the wave component for $K>1$.
For the system with N partons the relaxation time is scaled as $\tau^N \rightarrow N (\hbar/(mc\alpha)^2)$

# Introduction

In recent years, laser-plasma accelerators have gained more ground on becoming a promising alternative to conventional large scale particle acceleration facilities based on if technology. In laser generated wakefields, the electrons can experience accelerating fields of several hundred GV/m This exceeds the fields of conventional accelerators by several orders of magnitude. Thus, laser-plasma accelerators can generate high-energy particles over distances of only a few hundred micrometers. Electron energies of more than 300 MeV have been achieved in experiments .Generally, the generated electron beams exhibit a broad exponential spectrum and large divergence. Recent experiments have, however, demonstrated the capability of laser-plasma accelerators to produce quasi-monoenergetic and well-collimated electron bunches. This constitutes a major advance towards die application of laser-plasma accelerators to high-energy physics, biology, or medicine.

Scaling laws provide a very simple, even simplistic approach to understanding the very small systems: nuclei and nanoparticles. Detailed understanding requires sophisticated model dependent calculations. But basic scaling law calculations, used with appropriate care, can show why the very small systems have a very interesting properties common for example for nanoparticles and nuclei.

In this paper considering results published in our monograph [1], we investigate the scaling of the relaxation times for the nuclei and nanoparticles with N components- *partons.* It is shown that for the gas of N- *partons* the relaxation time for the system is scaled as $\tau^N \to N\tau$, where $\tau$ is the microscopic relaxation time $\tau = \hbar / m(\alpha c)^2$ and α is the coupling constant, α=1/137 for the electromagnetic interactions and 0.16 for strong interaction, c is the light vacuum speed

## Scaling law in quantum thermal phenomenal

In monograph [1] the quantum description of the quantum thermal phenomena was presented. It was shown that in the case of the ultra short laser pulses the heat transport is described by hyperbolic quantum heat transport equation [1]:

$$\tau \, \partial^2 T / \partial^2 t + \partial T / \partial t = D \Delta T \quad (2.1)$$

where *T* denotes the temperature of the *parton gas* in the system, $\tau$ is the relaxation time, *m* is the parton mass and *D* is the thermal diffusion coefficient. The relaxation time $\tau$ is defined as [1]:

$$\tau = h/mv^2 \qquad (2.2)$$

where v is the thermal pulse propagation speed:

$$v = a_i c \qquad (2.3)$$

In formula (2.3) $a_i$ is the coupling constant, $a_1 = 1/137$ for *electromagnetic interactions* and $a_2 = 0.15$ for *strong interactions*, c denotes the light velocity in vacuum. Both parameters $\tau$ and *v* completely characterize the thermal energy transport on the atomic scale and can be named as "microscopic" relaxation time and "microscopic" heat velocity.

In the following, starting with the microscopic $\tau$ and v we describe thermal relaxation processes in system( nanoparticle or nuclei) which consist of *N* partons. To that aim we use the Pauli-Heisenberg inequality [1]:

$$\Delta r \, \Delta p > N^{1/3} \, \hbar \qquad (2.4)$$

where *N* denotes the heat carriers, *partons*, in structure: *electrons* in nanoparticles or nucleons in the nuclei. From formula we conclude that for the system with N partons we can introduce the effective Heisenberg constant $\hbar^N$

$$\hbar^N = \hbar\, N^{1/3} \qquad (2.5)$$

According to formula (2.5) we recalculate the relaxation time $\tau$ and thermal velocity v for system consisting of *N partons*:

$$\hbar^N = \hbar\, N^{1/3} \qquad (2.5)$$

$$v^N = N^{-1/3}\, v \qquad (2.6)$$

$$\tau^N = N\, \tau \qquad (2.7)$$

Formula (2.7) describes the scaling for the relaxation time for the system of N *noninteracting partons*

Let us assume that system of N *partons* is heated and energy delivered to the system is equal E. As was shown in monograph [6] the response of the system depends on the value of the ratio K, viz.,

$$K_i = E/m_i(a_i c)^2 \qquad (2.8)$$

For K<1 the heating of the systems can be described as the thermal diffusion process. For K>1 the transport of the thermal energy contains the wave component [1] In Fig 1 the calculations of the K are presented for the electron gas and in Fig.2 for the nucleon gas. It is interesting to observe that for both systems the K=1 is crossed for the nearly the same value of energy scaled as $10^6$

In Fig.3 the parameter K is calculated as a function of the λ- wave length of the laser energy. It can be concluded that for λ < 300nm in electron gas the heat transport contains the wave component

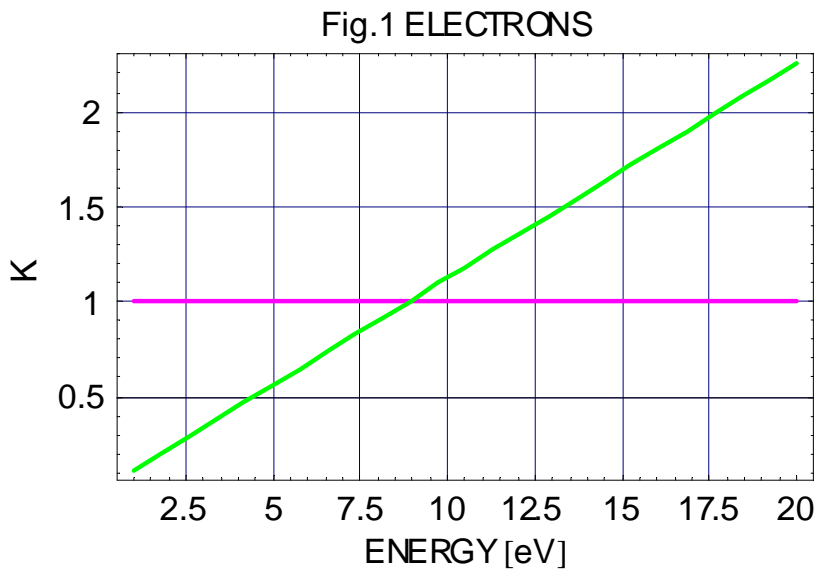

Fig.1 ELECTRONS

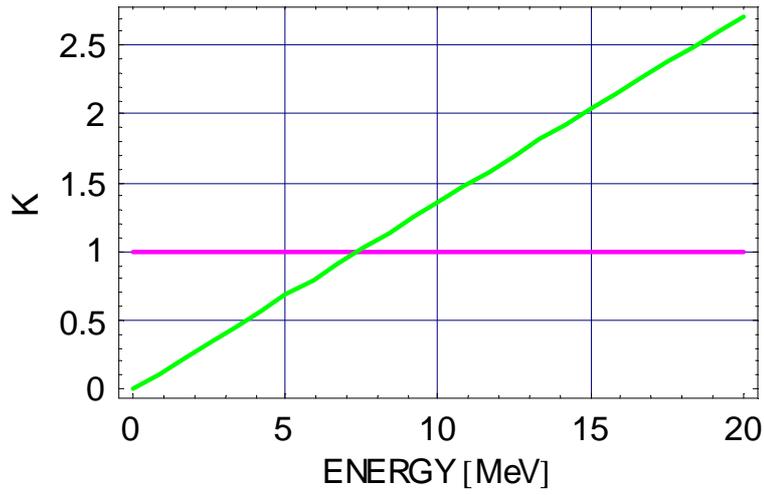

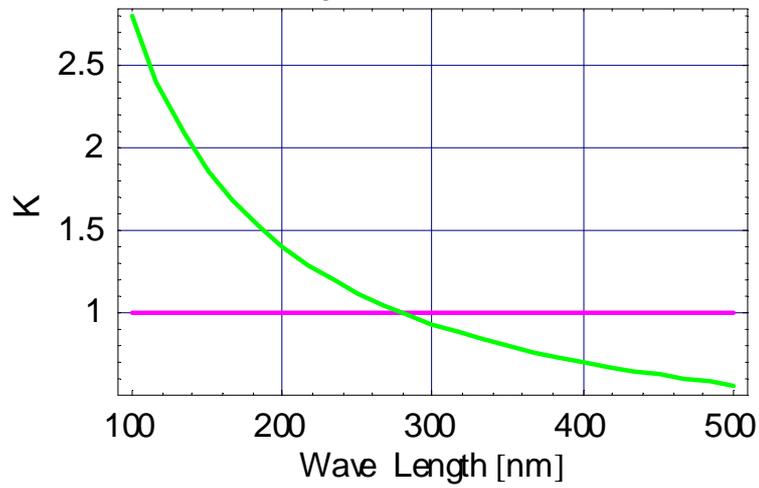

References

[1] M.Kozlowski and J. Marciak-Kozlowska, *Thermal phenomena using attosecond laser pulses*, Springer 2006